\documentclass[
twocolumn,
reprint,
superscriptaddress,
nofootinbib,
nobibnotes,
amsmath,
amssymb,
aps,
longbibliography,
prb,
showkeys
]{revtex4-2}

\usepackage[latin1]{inputenc}
\usepackage{amsmath}
\usepackage{amsfonts}
\usepackage{hyperref}
\usepackage{amssymb}
\usepackage{graphicx}
\usepackage{colortbl}
\usepackage{subfigure}
\usepackage{amsthm}
\usepackage{thmtools}
\usepackage[normalem]{ulem}
\usepackage{dsfont}
\usepackage{circuitikz}
\usepackage{cases}

\usepackage[T1]{fontenc}
\usepackage{mathptmx}
\usepackage{bbold}
\usepackage{scalerel}
\usepackage{xcolor}

\usepackage{mathtools}

\numberwithin{equation}{section}

\newcommand{\hyref}[1]{\hyperref[#1]{\ref{#1}}}

\newcommand{\orange}[1]

\renewcommand{\thesection}{\arabic{section}}
\renewcommand{\thesubsection}{.\arabic{subsection}}

\newcommand{\thenewsubsection}{\thesection.\arabic{subsection}}
\newcommand{\thenewsubsubsection}{\thesection\thesubsection.\arabic{subsubsection}}

\makeatletter
\def\@hangfrom@section#1#2#3{\@hangfrom{#1#2}#3}%\MakeTextUppercase{#3}}%
\def\@hangfroms@section#1#2{#1#2}%\MakeTextUppercase{#2}}%
\makeatother

\usepackage{titlesec}

\titleformat{\section}  % which section command to format
  {\fontsize{12}{14}\bfseries} % format for whole line
  {\thesection} % how to show number
  {1em} % space between number and text
  {\centering} % formatting for just the text
  [] % formatting for after the text

\titleformat{\subsection}  % which section command to format
  {\fontsize{10}{12}\bfseries} % format for whole line
  {\thenewsubsection} % how to show number
  {2em} % space between number and text
  {\centering} % formatting for just the text
  [] % formatting for after the text

\titleformat{\subsubsection}  % which section command to format
  {\fontsize{10}{12}\itshape} % format for whole line
  {\thenewsubsubsection} % how to show number
  {2em} % space between number and text
  {\centering} % formatting for just the text
  [] % formatting for after the text
\begin{document}

\title{Energy-efficient time series processing in real-time with fluidic iontronic memristor circuits}

\author{T. M. Kamsma}
\affiliation{Institute for Theoretical Physics, Utrecht University,  Princetonplein 5, 3584 CC Utrecht, The Netherlands}
\affiliation{Mathematical Institute, Utrecht University, Budapestlaan 6, 3584 CD Utrecht, The Netherlands}
\author{Y. Gu}
\affiliation{School of Aeronautics and Institute of Extreme Mechanics, Northwestern Polytechnical University, Xi'an 710072, China}
\author{C. Spitoni}
\affiliation{Mathematical Institute, Utrecht University, Budapestlaan 6, 3584 CD Utrecht, The Netherlands}
\author{M. Dijkstra}
\affiliation{Soft Condensed Matter \& Biophysics, Debye Institute for Nanomaterials Science,
Utrecht University, Princetonplein 1, 3584 CC Utrecht, The Netherlands}
\author{Y. Xie}
\affiliation{School of Aeronautics and Institute of Extreme Mechanics, Northwestern Polytechnical University, Xi'an 710072, China}
\author{R. van Roij}
\affiliation{Institute for Theoretical Physics, Utrecht University,  Princetonplein 5, 3584 CC Utrecht, The Netherlands}

\date{\today}

\begin{abstract}
Iontronic neuromorphic computing has emerged as a rapidly expanding paradigm. The arrival of angstrom-confined iontronic devices enables ultra-low power consumption with dynamics and memory timescales that intrinsically align well with signals of natural origin, a challenging combination for conventional (solid-state) neuromorphic materials. However, comparisons to earlier conventional substrates and evaluations of concrete application domains remain a challenge for iontronics. Here we propose a pathway toward iontronic circuits that can address established time series benchmark tasks, enabling performance comparisons and highlighting possible application domains for efficient real-time time series processing. We model a Kirchhoff-governed circuit with iontronic memristors as edges, while the dynamic internal voltages serve as output vector for a linear readout function, during which energy consumption is also logged. All these aspects are integrated into the open-source pyontronics package. Without requiring input encoding or virtual timing mechanisms, our simulations demonstrate prediction performance comparable to various earlier solid-state reservoirs, notably with an exceptionally low energy consumption of over 5 orders of magnitude lower. These results suggest a pathway of iontronic technologies for ultra-low-power real-time neuromorphic computation.
\end{abstract}

\maketitle

\section{Introduction}

The emerging field of iontronic neuromorphic computing has grown at an extraordinary rate in recent years \cite{Fan2025EmergingComputation,Lv2025AdvancementsDesign,SuwenLaw2025RecentComputing,Wang2026NeuromorphicConductors}. Consequently, exciting demonstrations of iontronic computing have been presented recently, including fluidic crossbar arrays \cite{Liu2025Resistance-RestorableChip}, (integrated) logic circuits \cite{Portillo2024ReversibleMemristor,Sabbagh2023DesigningCircuit,Han2023IontronicDissolution,Emmerich2024NanofluidicSwitches}, and reservoir computing \cite{Kamsma2024Brain-inspiredNanochannels,Armendarez2024Brain-InspiredPlasticity,Mohamed2024IntrinsicComputing,Mohamed2025Memimpedance-BasedComputing,Sarles2025Voltage-responsiveComputing,Portillo2025NeuromorphicDiodes,Li2025BiologicallyComputing}. These initial advancements increase the achieved functionality of iontronic circuitry for computing purposes, although they still remain at relatively simple levels for now. Consequently, power consumption estimates of fluidic circuits remain sparse as the focus is often on the power of individual device components. Moreover, power comparisons to other substrates on established tasks are challenging as many time-series benchmarks for comparisons remain out of reach. Thus, highlighting potential avenues for applications of iontronic neuromorphic devices remains difficult, especially when this ultimately has to be compared to the ultra-high performances of well-established solid-state technologies. Here we aim to both advance the theoretical guidance towards interacting iontronic circuits, including circuit-wide power estimates for performing established time-series benchmarks, and to advance the mapping out of potential domains of relevance for iontronic technologies.

\begin{figure*}[ht!]
    \centering
        \includegraphics[width=1\textwidth]{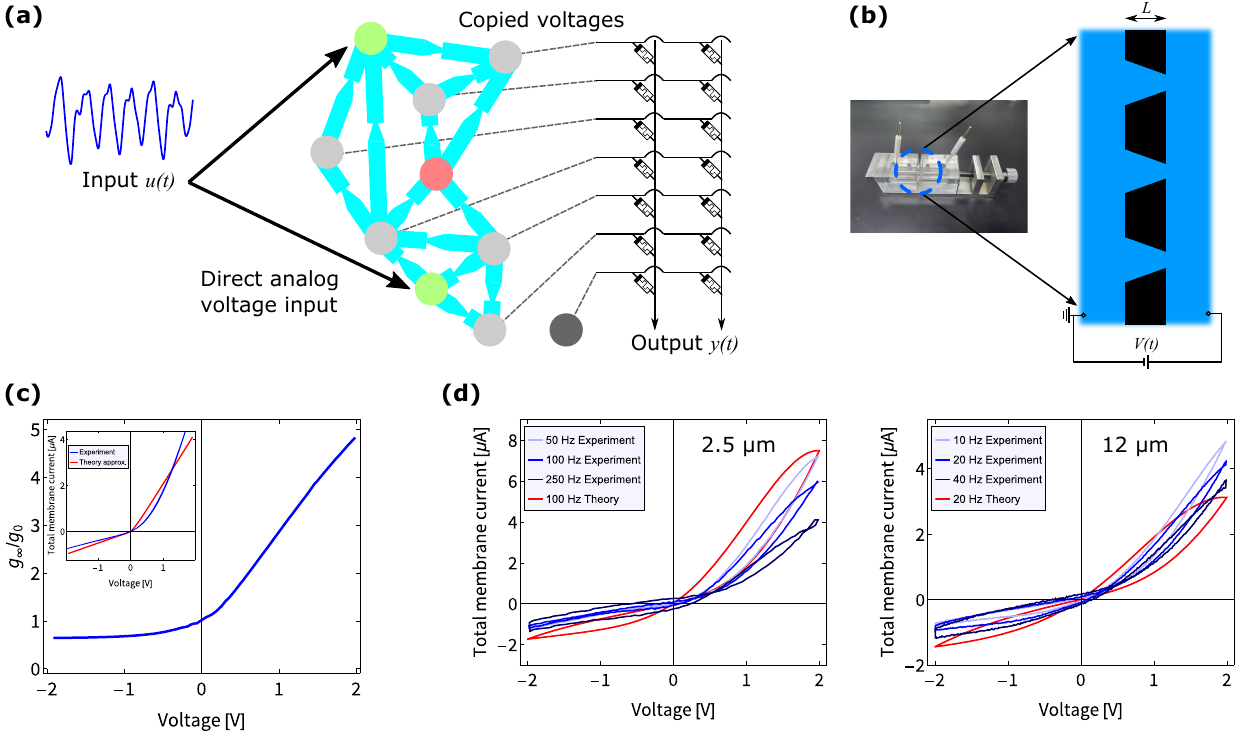}
    \caption{\label{fig:Fig1}\textbf{(a)} Schematic diagram of the modelled system, with a fluidic circuit consisting of iontronic memristors connected by aqueous reservoirs. A time series input is directly applied as imposed voltages at one or multiple of the reservoir nodes (green), with at least one ground (red). The free voltages in the remaining $N_{\text{free}}$ nodes (grey) evolve according to Kirchhoff's law and the dynamically evolving memristor conductances. These voltages, concatenated with constant 1 V offset voltage such that $\mathbf{V}_{\text{free}}(t)=(V_1,..,V_{N_{\text{free}}}, 1)$ V, are directly copied as voltage inputs into a linear readout function such that the output is $\textbf{y}(t)=W_{\text{out}}\mathbf{V}_{\text{free}}$. The physical implementation of such a matrix-vector multiplication is a standard crossbar array as depicted on the right. \textbf{(b)} Photograph of the full device that features a membrane connecting two aqueous electrolyte reservoirs. The membrane features a collection on conical channels that can conduct an ionic current. \textbf{(c)} Voltage-dependent steady state conductance as measured from the 12 $\mu\text{m}$ thick membrane. The inset shows the corresponding current-voltage relation from the experiment (blue) and a theoretical approximation (red) \cite{Boon2022Pressure-sensitiveGeometry}. \textbf{(d)} Dynamic $IV$ measurements (blue graphs) of membranes of thicknesses 2.5 $\mu\text{m}$ (left) and 12 $\mu\text{m}$ (right), revealing memristive hysteresis loops that emerge across different frequencies. In red are the dynamic currents as predicted by the SVM theory as in Eq.~(\ref{eq:dgdt}) with timescales $\tau$ of 1 ms and 10 ms for the short and long channels respectively.}
\end{figure*}
We highlight possible applications in efficient real-time processing of time series by exploiting the combined ingredients of (i) intrinsic alignment of the ms to s time scales of iontronic devices with those of signals of natural origin, (ii) iontronic dynamics that align well with established reservoir computing protocols \cite{Kamsma2025EchoSubstrate}, and (iii) the possibility that these features can scale down to angstrom-confined iontronic devices of ultra-low power consumption \cite{Shi2023UltralowMemristor}. Especially the first feature of natural timescales is notable, because achieving this within conventional materials proves challenging and many conventional neuromorphic circuits require additional hardware to artificially slow down its components or have to rely on a virtual time that is detached from real-time \cite{Chicca2020ASystems,Indiveri2025NeuromorphicNeuromorphic.}. This requires additional overhead and energy consumption. Iontronic memristors on the other hand can scale down to angstrom confined devices \cite{Shi2023UltralowMemristor,Biswabhusan2024AngstromStates,Robin2023Long-termChannels,Xu2025Angstrom-Scale-ChannelComputing,Liu2025Resistance-RestorableChip}, enabling low power consumption per channel, while retaining the relevant ms to s timescales \cite{Shi2023UltralowMemristor,Xu2025Angstrom-Scale-ChannelComputing,Liu2025Resistance-RestorableChip}. Therefore, energy-efficient real-time processing of time series of signals that evolve over slow natural timescales represents a possible area where iontronic technologies could prove to be relevant.

Various other domains of application have also been suggested for iontronics, with biological compatibility, ion selectivity, or chemical regulation often being named \cite{Fan2025EmergingComputation,Lv2025AdvancementsDesign,SuwenLaw2025RecentComputing,Wang2026NeuromorphicConductors}. While certainly exciting future directions, these suggestions often (for now) lack concreteness or rely on speculative technologies such as artificial fully ion-specific channels. Notably, to ensure that the proposed next steps here are concrete, we remark that the features of iontronic circuits modelled here have either already been experimentally demonstrated before, are being supported experimentally in this work, or are realistically feasible in the near future.

Below we will present the details of a modelled fluidic circuit containing 15 iontronic memristors. Our results are representative for a wide class of iontronic memristors, but we will also present measurements of specific iontronic memristive devices and base the conductance properties on these. The conductance equations of motion are incorporated in a fluidic network with the iontronic memristors forming the edges. The internal voltage dynamics in this circuit is then solved according to Kirchhoff's law, with the free voltages forming a reservoir computing network, so the physical properties of the fluidic circuit remain fixed. For the evaluation of the energy consumption we calculate the total internal Ohmic power dissipation of the (calculated) conductances and voltages in the iontronic circuit. All these calculations are smoothly integrated in a new addition to the \textit{pyontronics} software package \cite{pyontronics}. The free internal voltages are then used as vector input into a linear readout function for the task of predicting the (chaotic) Mackey-Glass time series \cite{Mackey1977OscillationSystems}. Therefore, the readout function is fully compatible with crossbar arrays as physical implementation, as these are a hardware realisations of matrix-vector operations and incorporate voltage vectors as direct input. Importantly, the input of the time series requires no additional pre-processing and the dynamic output voltages can be directly copied to be voltage inputs for the crossbar array readout function. Therefore, both the input and output peripheral circuitry can be kept at a minimum.

Our physical simulation of a fluidic circuit demonstrates a performance that is on par with earlier work using more established substrates, while operating at a power consumption that is over 5 orders of magnitude lower. Additionally, depending on the type of iontronic memristor, the circuit is able to directly analyse time series that have timescales of milliseconds to tens of seconds akin to many biological signals, without the need of input encoding or virtual clocks. This improvement in energy efficiency is driven by the low power consumption enabled by highly confined iontronic devices, with memristive angstrom scale channels featuring power consumptions as low as $\sim10$ fW per device \cite{Shi2023UltralowMemristor, Liu2025Resistance-RestorableChip}. Therefore, the advent of such iontronic memristors at the angstrom scale provides an exciting possible future perspective for ultra-low power time series processing in real-time.

\section{Circuit and device dynamics}
The fluidic circuit we study here is a straightforward network of iontronic memristors that obeys Kirchhoff's law, schematically depicted in (the middle of) Fig.~\ref{fig:Fig1}(a), which we will describe in more detail in Sec.~\ref{sec:sec2}. The iontronic memristive elements in this fluidic circuit are \textit{Simple Volatile Memristors} (SVMs) \cite{Kamsma2024ACircuits}. This class of volatile memristors is based on the natural assumption that the time-dependent conductance $g(t)$ between two nodes of the network, at voltage difference $V(t)$, satisfies  $\partial_{\mathrm{t}}g(t)=f\big(g_{\infty}(V(t))-g(t)\big)$ for some function $f$, with $g_{\infty}(V)$ the steady-state conductance. For stability arguments $f(0)=0$, so an expansion of $f$ up to first order yields the equation of motion of a leaky integrator conductance given by
\begin{align}
    \dfrac{\mathrm{d}g(t)}{\mathrm{d}t}=\frac{g_{\infty}(V(t))-g(t)}{\tau},\label{eq:dgdt}
\end{align}
where $\tau$ is the characteristic time scale (see below) on which the memristor responds to voltage changes. 
It has been demonstrated theoretically \cite{Kamsma2023IontronicMemristors,Kamsma2023UnveilingIontronics} and experimentally \cite{Kamsma2024Brain-inspiredNanochannels,Cervera2024ModelingDiodes} that various fluidic iontronic memristors behave as SVMs. Note that the above derivation implicitly assumes that there is only a single state variable that determines the conductance, in this case effectively the conductance $g(t)$ itself, determined by the dynamic salt concentration in transient ion concentration polarisation based memristors \cite{Kamsma2023IontronicMemristors}. Therefore, the presence of multiple conductance-determining physical processes with possible multiple timescales \cite{Robin2023Long-termChannels,Kamsma2025ChemicallyApplications,Jouveshomme2025MultipleMem-spectrometry,Rivera-Sierra2025RelaxationApplications,Cervera2025MultipleModel,Xu2025Angstrom-Scale-ChannelComputing} are not fully captured by the above framework, though individual conductance components can sometimes still be described through an SVM approach \cite{Robin2023Long-termChannels}. 

As the SVM description does seem to be a remarkably effective theoretical framework for various devices \cite{Markin2014AnPlants,Kamsma2023IontronicMemristors,Kamsma2023UnveilingIontronics, Zhang2024GeometricallySystems,Kamsma2024Brain-inspiredNanochannels,Cervera2024ModelingDiodes}, including devices that have demonstrated computing functionalities \cite{Sabbagh2023DesigningCircuit,Kamsma2024Brain-inspiredNanochannels,Portillo2025NeuromorphicDiodes}, we will incorporate SVMs in our fluidic circuit simulations. For various SVMs the timescale $\tau$ has been demonstrated depend on the channel length $L$ \cite{Kamsma2024Brain-inspiredNanochannels,Zhang2024GeometricallySystems}, with theoretical estimates $\tau\propto L^2/D$ \cite{Kamsma2023IontronicMemristors,Kamsma2024Brain-inspiredNanochannels} where $D\simeq 1\mu\text{m}^2/\text{ms}$ is the typical diffusion coefficient of microscopic ions in a (bulk) aqueous electrolyte. This is a crucial aspect as it allows for the inclusion of multiple timescales in a single circuit by varying the individual lengths of the SVMs. Moreover, different SVMs have been demonstrated to span orders of magnitude in timescales, with $\tau$ ranging from order $\sim 1-100$ ms \cite{Kamsma2023IontronicMemristors,Kamsma2023UnveilingIontronics,Kamsma2024AdvancedCircuit,Cervera2024ModelingDiodes} to $\sim 1-100$ s \cite{Kamsma2024Brain-inspiredNanochannels,Zhang2024GeometricallySystems}. Therefore, while the results in this work will be in the $\sim 1-10$ ms regime, all results should equivalently hold for slower inputs of e.g. $1-100$ s time scales (or anything in between), opening the window for an excitingly wide range of time series.

Although our results are in principle aimed to be representative for a wide class of iontronic memristors, we base the conductance properties here on experimentally realised iontronic memristors, consisting of conical ion channels embedded in a membrane. Here we do note that the presented results did not appear to be sensitive to the precise form of the steady-state conductance $g_{\infty}(V)$. Figure 1(b) shows the experimental setup. The conical nanochannels were fabricated within the PET (poly(ethylene terephthalate)) foils following the protocol of the track-etched technique \cite{apel2001nuclearinstrumentsandmethodsinphysicsresearchsectionb:beaminteractionswithmaterialsandatomsb,siwy2002phys.rev.lett.,siwy2003nuclearinstrumentsandmethodsinphysicsresearchsectionb:beaminteractionswithmaterialsandatomsa,sheng2017chem.commun.a,zhou2024natl.sci.rev.}. The PET foils were first irradiated by swift heavy ions in the Lanzhou Heavy Ion Research Facility (HIRFL), forming ion tracks along their trajectories. Then we placed the PET foils under UV exposure and followed by asymmetric chemical etching, which enables the fabrication of conical nanochannels along the ion tracks. The $2.5\;\mu\text{m}$ thick membrane has the track density $5\times 10^{9}\;\text{cm}^{-2}$ and $3.14\;\text{mm}^2$ conductive area exposed to electrolyte solutions, while $78.5\;\text{mm}^2 $ the conductive area and $ 3\times10^{8}\;\text{cm}^{-2}$ for the $12\;\mu\text{m}$ thick ones, for a comparable resistance at both sides. 
Two Ag/AgCl electrodes, connected to an electrochemical workstation (AMETEK ModuLab Xtreme) for triangular voltage sweeping, were inserted in reservoirs filled with neutral aqueous $0.1\;\text{M}$ KCl solution. The result is two membranes containing a large set of conical channels connecting aqueous electrolyte reservoirs, as schematically depicted on the right of Fig.~\ref{fig:Fig1}(b), with a thin membrane containing channels of length $L=2.5\;\mu\text{m}$ and a thick membrane containing channels of length $L=12\;\mu\text{m}$. 
Unlike the Angstrom-scale latent tracks fabricated by soft-etching, which has known numbers of tracks\cite{wen_highly_2016,wang_ultrafast_2018}, it is difficult to examine the exact numbers of nanochannels etched through by the track-etched technique\cite{siwy_fabrication_2002}. Below we will support a rough conductance estimate of $\sim 1$ pS per channel as an underestimation in this work, assuming all of the etched through channels feature a small $R_{\mathrm{t}}\sim 2$ nm tip that is slightly larger than the soft-etched angstrom channels that are $\sim 10$ fS per channel regime \cite{Shi2023UltralowMemristor,Liu2025Resistance-RestorableChip}.

From steady-state $IV$ measurements we can obtain the steady-state conductance $g_{\infty}(V)$, depicted in Fig.~\ref{fig:Fig1}(c). The interpolated function of these experimental results are used as $g_{\infty}(V)$ in Eq.~(\ref{eq:dgdt}). Due to the uncertainty in number of channels and polydispersity of the channels in the membrane, a precise estimate of conductance per channel is challenging, where the total current is likely dominated by the largest channels. Therefore we instead provide a rough estimate based on a theoretical model for current rectification in conical channels \cite{Boon2022Pressure-sensitiveGeometry}. The experimentally found current rectification is reasonably well in agreement with a theoretical approximation of channels of tip and base radii of $R_{\mathrm{t}}=2$ nm and $R_{\mathrm{b}}=8$ nm respectively, with other standard parameters of a diffusion coefficient of $D=1\text{ }\mu\text{m}^2\text{ms}^{-1}$ and surface charge density $e\sigma=-0.2\;e\text{nm}^{-2}$ (leading to a zeta potential of $\approx-40$ mV). The resulting steady-state $IV$ diagram is shown in the inset of Fig.~\ref{fig:Fig1}(c) in red for the long channel, compared to the experiments in blue. Channels of such high confinement feature equilibrium conductances of order $\sim 1$ pS, with these specific parameters $g_0=(\pi R_{\mathrm{t}} R_{\mathrm{b}}/L)(2\rho_{\rm{b}}e^2D/k_{\mathrm{B}}T)\approx 3.2$ pS, where $\rho_{\mathrm{b}}=0.1\;\text{M}$ is the salt concentration. With this in mind, we choose to set the equilibrium conductances $g_{i,0}=g_{i,\infty}(0)$ of the SVMs in our circuit to 1 pS. Here we stress that this is a conservative estimate in the broader scope of possible channels, with similar angstrom confined channels being demonstrated to feature two orders of magnitude lower conductances of $\sim 10$ fS \cite{Shi2023UltralowMemristor,Liu2025Resistance-RestorableChip}.

A crucial aspect of these iontronic memristors is that their conductance memory timescales $\tau_i$ are channel length $L_i$ dependent. To confirm this for the example SVMs presented here, and to confirm the dynamic conductance is captured by Eq.~(\ref{eq:dgdt}), we apply periodic triangle potentials of various frequencies to channels of length $2.5\;\mu$m and $12\;\mu$m. The resulting $IV$ diagrams are shown in Fig.~\ref{fig:Fig1}(d), with the short channels result on the left and the long channels results on the right. Here we not only see that the theoretically predicted pinched hysteresis loop is indeed found experimentally, we also find that these occur for frequencies separated by an order of magnitude, with the short channels featuring $\tau_i\sim 1$ ms and the long $\tau_i\sim 10$ ms. The precise values of $\tau_i$ are not crucial in this regard, as the circuit will feature a spread of them and we found that the precise values within this spread did not affect performance significantly. That there is at least some heterogeneity in the timescales is of importance though. Therefore, by varying the lengths of the different SVMs, such a spread in timescales can be achieved.

\subsection{Fluidic memristor network}\label{sec:sec2}
Consider a Kirchhoff-governed fluidic network with SVMs as edges and aqueous reservoirs as nodes, an example of which is schematically depicted in Fig.~\ref{fig:Fig1}(a). The SVM edges feature dynamic conductances $\mathbf{g}(t)\in\mathds{R}^{M}$, with $M$ the number of SVMs. The aqueous reservoirs are at dynamic voltages $\mathbf{V}(t)\in\mathds{R}^N$ with $N$ the number of nodes in the network. A time series input $u(t)\in\mathds{R}$ can be directly applied as a voltage at one or multiple of the nodes with an optional different scaling prefactor or offset per node, but without any further input encoding. Lastly, at least one node is grounded. Therefore, the circuit includes $N_{\text{free}}=N-N_{\text{imp}}$ free voltages, with $N_{\text{imp}}$ the number of nodes where a voltage is imposed (including ground). Using Kirchhoff's law for given conductances $\mathbf{g}(t)$, the voltages at time $t$ are numerically calculated, after which the dynamic conductances are calculated for the next time step $t+\Delta t$ using Eq.(\ref{eq:dgdt}) and a straightforward Euler Method $\mathbf{g}(t+\Delta t)=\mathbf{g}(t)+\Delta t\partial_{t}\mathbf{g}(t)$, starting from initial conditions $\mathbf{g}(0)=\mathbf{g}_{\infty}(0)$. Therefore it is of importance that $\Delta t\ll \text{min}\{\tau_1,\cdots,\tau_M\}$, with $\tau_i$ the characteristic time scale of the $i$-th SVM in the circuit.

The dynamically evolving free voltages at the nodes of a given network are concatenated with a constant voltage of an (arbitrary) value 1 V to serve as a bias term, such that the total voltage vector is $\mathbf{V}_{\text{free}}(t)=(V_1,...,V_{N_{\text{free}}},1)$ V. The resulting voltage vector $\mathbf{V}_{\text{free}}(t)$ serves as a direct input for a linear readout function to obtain the output $\mathbf{y}(t)=\mathbf{W}_{\text{out}}\mathbf{V}_{\text{free}}(t)$, where $\mathbf{W}_{\text{out}}\in\mathds{R}^{O\times (N_{\text{free}}+1)}$ is the output matrix that is trained  using ridge regression \cite{Hoerl1970RidgeProblems} to reproduce a desired output $\tilde{\mathbf{y}}(t)\in\mathds{R}^{O}$. In this work $O=1$, and the desired output is chosen to be a prediction of the future input signal $\tilde{\mathbf{y}}(t)=u(t+\tilde{t})$ for a certain prediction window $\tilde{t}>0$. Standard crossbar arrays are the hardware equivalent of the required matrix multiplication and directly incorporate voltages as input vector. Therefore, besides input and output components for the voltage input and current output, the only required peripheral circuitry is a connection that copies voltages from the dynamic circuit to the input voltages for the standard crossbar array, as also depicted in Fig.~\ref{fig:Fig1}(a). In this work $\mathbf{y}(t)\in\mathds{R}^1$, but two voltage lines are drawn to represent the option of negative elements in $\mathbf{W}_{\text{out}}$, with one voltage line corresponding to positive matrix elements and the other to negative elements. The power consumption of the crossbar array is not taken into account when we calculate the power consumption. While this would contribute to the eventual total power consumption, it can be ignored when comparing the different reservoir computing implementations as we do here as all reservoir computing protocols require a readout function.

We note that the dynamic conductances rather than the voltages can also serve as output vector (in which case $\mathbf{W}_{\text{out}}\in\mathds{R}^{O\times M+1}$). This comes with the benefit of increased performance, in the sense that there will typically be more (conductance) edges in a network than free (voltage) nodes. However, we expect free voltages to be significantly easier to measure, especially at exceedingly low currents. Moreover, this approach requires little additional hardware, as the voltages only need to be copied to serve as voltage input for the readout crossbar array. In contrast, using the conductances would require an intermediate step of measuring currents and voltages to calculate conductances, followed by an additional step to convert these conductances back to voltages for input of the crossbar array. Within the accompanying code \cite{pyontronics}, choosing the conductances or voltages as output is one straightforward setting. As the aim of this work is to provide a concrete and realisable circuit, and a more complete energy consumption estimate, we will focus on the free voltages as a feature vector for the readout function.

\begin{figure*}[ht!]
    \centering
        \includegraphics[width=1\textwidth]{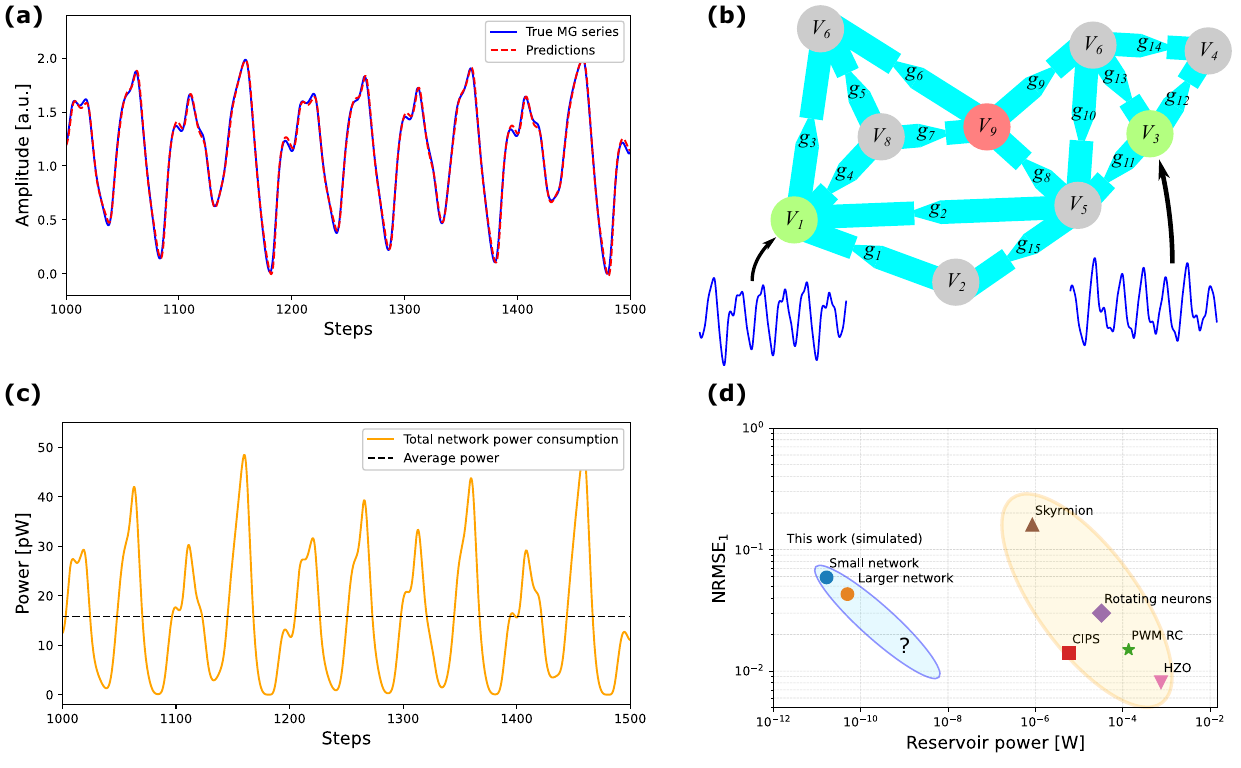}
    \caption{\label{fig:Fig2}\textbf{(a)} Mackey-Glass series (blue) and the predictions for 1 step ahead from a physical fluidic circuit containing 15 iontronic memristors (red), with step size 1 ms. This yielded a $\text{NRMSE}_{1}\approx0.059$, averaged over 4900 steps on a test series. \textbf{(b)} Schematic depiction of the fluidic circuit with green and red the input and ground nodes respectively, while the remaining grey nodes represent free voltages. \textbf{(c)} Total power usage of the fluidic circuit at any given time (orange) and on average (black), using the estimate of an individual channel equilibrium conductance of 1 pS. \textbf{(d)} Comparisons of the power consumption of the reservoir circuit compared to various literature results that report the same MG $\text{NRMSE}_{1}$ performance metric \cite{Liang2022RotatingComputing,Sun2023ExperimentalSystem,Liu2023Interface-typeComputing,Wu2024ACuInP2S6,Ma2025VersatileCircuits}. The blue circle corresponds to the small 15 SVM network discussed in Figs.~(a-c) and the orange circle represents the larger network of 3 parallel implementations of the circuit in (b) with $\text{NRMSE}_{1}\approx0.043$. A power consumption improvement of over 5 orders of magnitude is visible when compared to rotating neurons \cite{Liang2022RotatingComputing}, skyrmion-enhanced strain-mediated devices (Skyrmion) \cite{Sun2023ExperimentalSystem}, $\text{Hf}_{0.5}\text{Zr}_{0.5}\text{O}_2$ (HZO) transistors \cite{Liu2023Interface-typeComputing}, $\text{CuInP}_2\text{S}_6$ (CIPS) memristors \cite{Wu2024ACuInP2S6}, and pulse-width modulation resistor-capacitor (PWM RC) \cite{Ma2025VersatileCircuits}. An increased performance is found for higher power consumption in the literature results (yellow oval), also considerably driven by the differing reservoir sizes. Going from 1 circuit of SVMs (blue circle) to 3 parallel implementations of this circuit (orange circle) also improves performance, but whether such a trend continues (blue oval) remains to be investigated.}
\end{figure*}

\section{Efficient real-time reservoir computing}
The input signal we will use here is the Mackey-Glass (MG) series \cite{Mackey1977OscillationSystems}, which for right parameters forms a chaotic attractor \cite{DoyneFarmer1982ChaoticSystem}, as illustrated in Fig.~\ref{fig:Fig2}(a) in blue. The MG series is a ubiquitous synthetic time series for the benchmarking recurrent neural network frameworks \cite{Soltani2023EchoReview} and is also commonly used for implementations of hardware-based reservoir computing \cite{Moon2019TemporalSystem,Liu2024NeuromorphicApplications,Liang2022RotatingComputing,Sun2023ExperimentalSystem,Liu2023Interface-typeComputing,Wu2024ACuInP2S6,Ma2025VersatileCircuits}. Therefore, the MG series enables comparisons of (simulated) iontronic reservoir computing with earlier hardware-based time series prediction tasks.

We will focus on a fluidic network topology comprising of a total of 15 SVM edges, with 2 imposed voltage nodes and 1 ground node, schematically depicted in Fig.~\ref{fig:Fig2}(b). The network topology is inspired by the circuit layout used in Ref.~\cite{Dillavou2022DemonstrationLearning}, which appeared to consistently provide good performance compared to other configurations. While various different topologies yielded similar prediction results, the precise connections and orientations do affect the network's performance. Scanning such different topologies is straightforward through the simulation software \cite{pyontronics} and could therefore assist with experimental design. The network naturally serves as a reservoir computer where the dynamical states, either memristor conductances or node voltages, provide a high-dimensional representation of the input signal. Here we focus on the node voltages due to the aforementioned more straightforward hardware implementation.

The MG series is generated by
\begin{align}\label{eq:MG}
    \dfrac{\text{d}P(t)}{\text{d}t}=\frac{\beta P(t-t_{\text{delay}})}{\theta+P(t-t_{\text{delay}})^{n}}-\gamma P(t),
\end{align}
where we use $\beta=0.2$, $\theta=1$, $\gamma=0.1$, $n=10$, and $t_{\text{delay}}=17$ (we comment on the time units below), for which Eq.~(\ref{eq:MG}) is known to feature a chaotic attractor \cite{DoyneFarmer1982ChaoticSystem}. The first 17 time steps are randomly generated values in the range $[-1,1)$.

For in-silico predictions, the time $t$ is typically dimensionless, where ``1 step'' $\delta$ corresponds to one  dimensionless time step of $t\rightarrow t+1$. Similar to Ref.~\cite{Jaeger2010The1} we integrate the MG series $P(t+\Delta t)=P(t)+\Delta t\partial_tP(t)$ with steps $\Delta t=0.1$, so ``1 step'' of $\delta=10\Delta t$ corresponds to $10$ Newton steps. This yields a discrete sequence of values $P(n\delta)$, with $n\in\mathds{N}$, where the first natural prediction to make is the next step \cite{Liang2022RotatingComputing,Sun2023ExperimentalSystem,Liu2023Interface-typeComputing,Wu2024ACuInP2S6,Ma2025VersatileCircuits}, with 6 steps and 84 steps also being common for more complex (software) analysis methods \cite{Gers2001ApplyingApproaches}. One can choose what units these time steps represent and since iontronic SVMs can span timescales ranging from 1 ms to 100 s \cite{Kamsma2023IontronicMemristors,Kamsma2023UnveilingIontronics,Kamsma2024AdvancedCircuit,Cervera2024ModelingDiodes,Kamsma2024Brain-inspiredNanochannels,Zhang2024GeometricallySystems} a large range of possible different signal timescales could be analysed. Since our example SVMs feature timescales of 1-10 ms, the dimension of the MG time we consider in the specific physical simulations here is ms, thus $\delta=1$ ms. However, for SVMs with timescales of e.g.\ 1-10 s \cite{Kamsma2024Brain-inspiredNanochannels,Zhang2024GeometricallySystems}, equivalent simulations, and thereby identical performance, could be achieved by implementing the same $\tau_i$, but now with units seconds, in which case a step $\delta$ would be 1 s. To reflect this flexibility on timescales, we will present the MG time series results in dimensionless ``steps'', rather than dimensional time, shown Fig.~\ref{fig:Fig2}(a) in blue, with the notion that a step could represent 1 ms up to 1 s, depending on the type of memristors included in the circuit.

The performance metric we use is the normalised root mean squared error (NRMSE) of the prediction $\hat{P}_i(t+\tilde{t})$ with the true value $P_i(t+\tilde{t})$ for some given future time $t+\tilde{t}$, with $\tilde{t}>0$. Here we calculate the NRMSE for a specific $\tilde{t}$ according to
\begin{align*}\text{NRMSE}_{\tilde{t}}=\sqrt{\sum_{i=1}^{T}\frac{\big(\hat{P}_i(t+\tilde{t})-P_i(t+\tilde{t})\big)^2}{\sigma^2T}}, 
\end{align*}
with $T$ the number of steps (i.e. the period) over which the predictions are measured and $\sigma^2$ the variance of the true signal. We note that $T$ must be sufficiently large to guarantee that the NRMSE is independent of the specific measurement interval to ensure we do not coincidentally measure a period that is easier to predict, as some periods in the MG series are more chaotic than others. Various values of $\tilde{t}$ are commonly used, here we will focus on a single ``step'' ahead, i.e.\ $\tilde{t}=\delta$. As mentioned above, with SVMs of timescales 1-10 ms this corresponds to $\tilde{t}=1$ ms, while for slower inputs and slower SVMs a single prediction step could correspond to $\tilde{t}=1$ s. Such a single step ahead is a typical problem for hardware based implementations \cite{Liang2022RotatingComputing,Sun2023ExperimentalSystem,Liu2023Interface-typeComputing,Wu2024ACuInP2S6,Ma2025VersatileCircuits}, while 6 steps and 84 steps are also common in software based implementations \cite{Gers2001ApplyingApproaches}.

To ensure the simulation corresponds to a hardware implementation that is as straightforward as possible, we restrict ourselves to little to no pre-processing of the MG input. The only transformation we employ is a linear mapping of the dimensionless Eq.~(\ref{eq:MG}) to the voltage interval $\left[0,2\right]$ V, yielding our input $u(t)$. We did find that the precise voltage interval does not significantly impact the performance. The inputs $u(t)$ and $-u(t)$ are then applied as imposed voltages to nodes 1 and 3 respectively, while node 9 is ground, as depicted in Fig.~\ref{fig:Fig2}(b). Since our example SVMs presented in Fig.~\ref{fig:Fig1} exhibit timescales on the order 1-10 ms, we set the timescales of the SVMs with conductances $\left(g_1(t),...,g_M(t)\right)$ to $\left(\tau_1,...,\tau_M\right)=\left(1,1.5,2,...,8.5\right)$ ms, which thus would correspond to channels of length between $\sim2.5\;\mu\text{m}$ and $\sim12\;\mu\text{m}$ for our example SVMs. While heterogeneity in the timescales was of importance, we note that the performance did not appear to be sensitive to the precise distribution of timescales $\mathbf{\tau}$. Therefore, this suggests that solely incorporating timescale heterogeneity through varying lengths is sufficient, while no precise engineering is required to achieve specific timescales. The 6 free voltages $\mathbf{V}(t)$, together with the constant 1 V offset voltage, form the 7-dimensional dynamic feature vector such that $\hat{P}(t+\tilde{t})=\mathbf{W}_{\text{out}}\mathbf{V}(t)$, with $\tilde{t}=1$ ms. The linear readout function $\mathbf{W}_{\text{out}}\in\mathds{R}^{1\times 7}$ is trained using ridge regression \cite{Hoerl1970RidgeProblems} with a regularisation of $1\cdot10^{-4}$. The resulting predictions are shown in red in Fig.~\ref{fig:Fig2}(a). Measured over a total time interval of [0.1,5] s (i.e.\ 4900 steps), a performance of $\text{NRMSE}_{1}\approx0.059$ was achieved. An implementation of 3 networks as depicted in Fig.~\ref{fig:Fig2}(c) in parallel (i.e.\ 45 SVMs), but with differing timescale distributions, input, and ground nodes (full details can be found in accompanying code \cite{pyontronics}), improved the performance to $\text{NRMSE}_{1}\approx0.043$.

\subsection{Power usage estimate and comparisons}
The power the circuit consumes at any time $t$ is straightforwardly calculated as $\sum_{i=1}^{M}I_i(t)V_i(t)=\sum_{i=1}^{M}g_{i}(t)V_i(t)^2$. For angstrom scale channels $g_{i}(t)$ can be remarkably small, down to $\sim 10$ fS per channel \cite{Shi2023UltralowMemristor}. As discussed in Sec.~\ref{sec:sec2}, there is some uncertainty in the precise conductance per channel for the example SVMs we presented in Fig.~\ref{fig:Fig1}. With the conservative estimate of $g_{i,0}=1$ pS per channel we find the power usage results shown in Fig.~\ref{fig:Fig2}(c) in orange. Here we reiterate that in this specific example 1 step is 1 ms, while for SVMs of slower timescales \cite{Kamsma2024Brain-inspiredNanochannels,Zhang2024GeometricallySystems} the same results would be expected to emerge with 1 step representing e.g.\ 1 s. While the time units are of importance for total energy consumption, we focus on power usage here due to the ``always-on'' nature of such real-time time series analysis hardware, as we will elaborate more on below. In Fig.~\ref{fig:Fig2}(c) we see that the 15 SVMs of $\sim 1$ pS per channel, driven by $\sim 1$ V potentials, yields total reservoir circuit power estimates of $\sim10$ pW, with an average power consumption of $\overline{P}=15.8$ pW.

Comparisons of the power consumption of reservoir circuits can be remarkably deceptive for various reasons. First, power or energy usage is often not explicitly reported and hence needs to be inferred from other results. Secondly, for the real-time analysis of time series a circuit needs to be essentially ``always on'' and there typically is not a natural way to outline an energy per discrete task as a time series is continuously ongoing. Therefore, we choose to report a power rather than an energy metric. The time series is also frequently pre-processed or encoded by additional hardware \cite{Liang2022RotatingComputing,Liu2023Interface-typeComputing,Wu2024ACuInP2S6,Ma2025VersatileCircuits} as e.g.\ voltage pulses and the typical ``energy per pulse'' can then be reported \cite{Ma2025VersatileCircuits}. This energy metric is often reported in neuromorphic literature \cite{Xia2025Low-PowerApplications}, however its use in always-on circuits is rather indirect as the actual power consumption now relies on the sparsity of pulses. Additionally, this offloads a considerable part of the always-on and processing hardware, and thereby power consumption, to external components. Therefore, even if some energy metric is provided, this is not always directly applicable for comparisons.

With the above considerations in mind, we do estimate the power consumption of 5 earlier literature hardware implementations that report the $\text{NRMSE}_1$ metric on the MG series \cite{Liang2022RotatingComputing,Sun2023ExperimentalSystem,Liu2023Interface-typeComputing,Wu2024ACuInP2S6,Ma2025VersatileCircuits}. We do note that this is not necessarily an exhaustive list and the works that specifically report the MG $\text{NRMSE}_1$ form only a small subset of all analyses of time series carried out directly in hardware. Combining the performance and (estimated) power usage yields the results in Fig.~\ref{fig:Fig2}(d), where we see over 5 orders of magnitude improvements in power usage for the reservoir circuit. The predominant driving factor behind this power improvement is the exceptionally low power consumption per channel, enabled by highly confined channel geometries. We reiterate that the precise power usage values feature some uncertainty (except for Ref.~\cite{Liang2022RotatingComputing} which does provide an explicit estimate). However, the differences in power consumption span several orders of magnitude and the estimates will rather be lower bounds than upper bounds, as we ignore peripheral circuitry energy consumption.

The power usage of peripheral hardware could potentially be minimal for our simulated system as little to no input or output encoding is needed for our circuit design. Firstly, the input can be directly linearly mapped to a voltage. This analog input means that in principle no analog-to-digital (ADC) conversion is needed, where ADCs in earlier circuits could consume up to 80-88\% of the chip's energy \cite{Aguirre2024HardwareNetworks}. Moreover, since the outputs are also voltages, these can directly be copied into the voltage input of a crossbar array that could serve as the linear readout function. Therefore, additional power-consuming peripheral circuitry could possibly be kept at a minimum.

In Fig.~\ref{fig:Fig2}(d) there is a clear trend visible in the literature results, with higher power leading to better performance (yellow oval). The larger network of 3 parallel circuits consumed $\overline{P}=39.8$ fW and improved the $\text{NRMSE}_{1}$ from $\approx0.059$ to $\approx0.043$. While we thus see an improvement when we triple the SVMs (and thus the power consumption), it remains to be seen whether that trend (blue oval) extrapolates to even larger networks. It could be that the introduction of some peripheral circuitry could help the improvement of the network further at the cost of additional power consumption, e.g.\ it has been shown that, with additional supporting hardware, iontronic networks can be full hardware equivalents of established echo state and band-pass networks \cite{Kamsma2025EchoSubstrate}. Whether the potentially improved performance is worth the additional power cost of such extra hardware represents an interesting and relevant trade-off to pursue in further work.

% \subsection{Network improvements over parallel implementation}
% Integrating iontronic devices remains a challenge, and our circuit of 15 iontronic memristors would be already near the current fabrication capabilities \cite{EdriFraiman2025TowardIntegration}. A connected network topology complicates experimentally realisability further, which raises the question whether the same devices connected in parallel might perform equally well. Firstly, such a parallel configuration, while simplifying fabrication, would complicate the actual inference because there are no free voltages. Therefore, the feature vector must be constructed using the more involved method of measuring currents and calculating conductances. Moreover, this approach also complicates the ultra-low power consumption, as unlike free voltages, ultra-low currents are extremely difficult to reliably measure.

% To also demonstrate that a connected network exhibits an improved performance, we compare the performance of the network to the same SVMs in a parallel configuration, schematically depicted in Fig.~\textcolor{red}{TO DO}, in both instances using the dynamics conductances as feature vector. 

\section{Discussion and Conclusion}
In summary, we have presented a simulated fluidic circuit, containing iontronic memristors, and predicted a comparable performance to various previous hardware based implementations on predicting the common Mackey-Glass time series benchmark at an estimated 5 orders of magnitude power consumption improvement of the reservoir circuit \cite{Liang2022RotatingComputing,Sun2023ExperimentalSystem,Liu2023Interface-typeComputing,Wu2024ACuInP2S6,Ma2025VersatileCircuits}. The free voltages in the reservoir nodes form the dynamic output, which are fed into a linear readout function. This is fully compatible with crossbar arrays as a hardware readout function implementation. Since a circuit needs to be essentially ``always on'' for the real-time analysis of time series and since a time series is typically continuously ongoing, we focus on power as energy metric for this specific context. Iontronic memristors scale down to angstrom confined channels \cite{Shi2023UltralowMemristor,Biswabhusan2024AngstromStates,Robin2023Long-termChannels,Xu2025Angstrom-Scale-ChannelComputing,Liu2025Resistance-RestorableChip}, consequently the power consumption per channel can be exceedingly low, with earlier demonstrated devices going as low as $\sim 10$ fW per channel upon $\sim1$ V inputs \cite{Shi2023UltralowMemristor,Liu2025Resistance-RestorableChip}. Here we use a rather conservative estimate of $\sim 1$ pW per channel. Aiding the feasibility of our work is that angstrom confinement is not a requirement, but rather an optimal case for the best energy efficiency. For such optimal efficiency, the engineering and characterisation of angstrom-scale channels is crucial, where also interesting understandings are to be gained about the inner working of such highly confined systems \cite{Fong2024TheElectrolytes,ONeill2024ToWater,Fong2025OnElectrolytes} which might offer additional exploitable dynamics \cite{Robin2023Long-termChannels,Xu2025Angstrom-Scale-ChannelComputing}. With the conservative estimate of a conductance of order $\sim 1$ pS per channel, the full fluidic reservoir circuit carries out its prediction task with an average total power consumption of $\sim 16$ pW, which is over 5 orders of magnitude less than various previous hardware based implementations \cite{Liang2022RotatingComputing,Sun2023ExperimentalSystem,Liu2023Interface-typeComputing,Wu2024ACuInP2S6,Ma2025VersatileCircuits}.

While the above result promises to be a considerable improvement in power consumption, the used comparisons \cite{Liang2022RotatingComputing,Sun2023ExperimentalSystem,Liu2023Interface-typeComputing,Wu2024ACuInP2S6,Ma2025VersatileCircuits} do not form an exhaustive list of hardware results that were applied to predict the Mackey-Glass time series. Moreover, the results that look at the same Mackey-Glass time series form only a subset of all results on the hardware that can process time series. Therefore it is entirely possible that other hardware-based implementations of time series analysis could narrow this power consumption gap. Therefore, expanding such a fluidic circuit to leverage the chemical \cite{Han2023IontronicDissolution,Robin2023Long-termChannels,Wang2024AqueousMembranes,Xiong2023NeuromorphicMemristor,Ling2024Single-PoreCombinations,Kamsma2025ChemicallyApplications} or pressure responsiveness \cite{Jubin2018DramaticNanopores,Barnaveli2024Pressure-GatedProcessing,Kamsma2025EchoSubstrate,conte2025multimodal} of fluidic circuits could further strengthen a potential domain of relevance for iontronics. Nevertheless, the reduction over 5 orders magnitude down to a power consumption of only $\sim 16$ pW for the full reservoir circuit in and of itself also does suggest an exciting possible direction for iontronic dynamic circuits.

Conventionally, the memristor conductance is often used as dynamic output function of the reservoir, but this poses challenges as the low conductance per channel implies similarly low currents at typical $\sim 1$ V potentials. To overcome this, we restricted ourselves to using the dynamic free voltages in the reservoir nodes, rather than the dynamic voltages. This also features the added benefit of a minimum of required output peripheral circuitry, as the internal circuit voltages can directly be copied as the voltage inputs for potential crossbar arrays that serve as the linear readout function. Moreover, the time series is directly applied as dynamic input, with no pre-processing besides a linear mapping of the input to a voltage interval. Therefore, also the input peripheral circuitry could be kept at a minimum.

While the above-mentioned promising results are based on experimentally feasible memristor circuit characteristics, they are (mostly) based on simulations and therefore remain predictions for now. Circuits of the type modelled here have not yet been presented, however integrated fluidic circuits of a similar number of device components have been achieved \cite{Sabbagh2023DesigningCircuit,EdriFraiman2025TowardIntegration,Liu2025Resistance-RestorableChip}. In parallel with this, integrating individually engineered angstrom channels is also challenging, although channels of such high confinement can be fabricated \cite{Shi2023UltralowMemristor,Xu2025Angstrom-Scale-ChannelComputing} and membranes with such channels have been successfully integrated before in crossbar arrays \cite{Liu2025Resistance-RestorableChip}. Moreover, no additional complicated (manipulations of) hardware is assumed, as the circuit is fixed, input is directly applied as a voltage and the output voltages can be directly copied into a standard crossbar array. Therefore, the modelling here is carefully constructed to represent an implementation that, while for now not yet achieved, might be realistically feasible in the near future. This is in line with one of the intents of this work to explicitly propose a future direction for the field to exploit the energy-efficient iontronic dynamics that align well with the longer ms to s timescales of signals of natural origin. Therefore, this work aims to advance both the theoretical guidance of neuromorphic iontronics toward more advanced (time series) computing tasks and the evaluation of potential domains of relevance for iontronic technologies.

\begin{acknowledgments}
Y.X. and Y.W. would like to acknowledge their funding NSFC 12388101 and 12241201.
\end{acknowledgments}

%\bibliography{references,bibfile}


\end{document}